\documentclass[aps,twocolumn,showpacs, amsmath,amssymb]{revtex4} 

\usepackage{dcolumn}
\usepackage{bm}
\usepackage{graphicx}
\usepackage{color}
\DeclareGraphicsExtensions{. jpg, . pdf, . mps, . png, . eps, . ps, . EPS}
\begin{document}
\def\be{\begin{equation}}
\def\ee{\end{equation}}
\def\bc{\begin{center}} 
\def\ec{\end{center}}
\def\bea{\begin{eqnarray}}
\def\eea{\end{eqnarray}}
\newcommand{\avg}[1]{\langle{#1}\rangle}
\newcommand{\Avg}[1]{\left\langle{#1}\right\rangle}

\title{Bose-Einstein distribution, condensation transition and multiple stationary states\\
in multiloci evolution of diploid populations}
\author{Ginestra Bianconi$^1$ and Olaf Rotzschke$^2$}
\affiliation{$^1$Physics Department, Northeastern University, Boston, Massachusetts 02115 USA, \\$^2$Singapore Immunology Network (SIgN), Biomedical Sciences Institutes (BMSI), Agency for Science, Technology and Research (A*STAR), IMMUNOS, 138648, Singapore,Singapore}
\begin{abstract}
The mapping between genotype and phenotype is encoded in the complex web of epistatic interaction between genetic loci. In this rugged fitness landscape, recombination processes, which tend to increase variation in the population, compete with  selection processes that tend to reduce genetic variation. Here we show that  the Bose-Einstein distribution  describe the multiple stationary states of a diploid population under  this multi-loci evolutionary dynamics. 
Moreover,  the evolutionary process might undergo an interesting condensation phase transition in the universality class of a   Bose-Einstein condensation when a finite fraction of pairs of linked loci, is fixed into given allelic states. Below this phase transition  the genetic variation  within a species  is significantly  reduced and only maintained  by the remaining polymorphic loci.
\end{abstract}
\maketitle
\section{Introduction}

The deep relation between evolutionary theory and statistical mechanics has been fascinating most of the scientists working in the field.  Historically,  Fisher  compared his fundamental theorem of natural selection to the second law of thermodynamics \cite{Fisher} and  Kimura, referring to the stochasticity of the evolutionary process, has compared  the genetic theory of evolution to  the theory of gases \cite{Kimura}. More recently,  further relations between evolutionary theory and statistical mechanics have been identified \cite{Hirsh,Gerland}. Indeed the relation between {\it chance and necessity} \cite{Monod}
in evolutionary theory, i.e.  the  trade-off between stochastic processes and selection, has the potential to be fully described  in statistical mechanics terms.
Interestingly, the relation between evolutionary theory and quantum statistical mechanics is emerging  from  a series of independent works \cite{Eigen,Peliti,Nowak,Kingman,Bose,Kadanoff,Ferretti,Shraiman} where it has been highlighted that a class of  phase transitions occurring in evolution of haploid population can be described in terms of a Bose-Einstein condensation. These transitions are called in the biological literature quasispecies transitions and represent collective change in the populations when mutations compete with natural selection. A quasispecies transition that can be mapped to a Bose-Einstein condensation,  occurs in asexual populations when the mutation rate is below a  critical value where   a finite fraction of the population is localized in sequence space around a given genotype.
 In this paper we    show that the Bose-Einstein distribution and condensation transition in the Bose-Einstein universality class  occur  in the evolutionary theory of diploid populations when recombination competes with selection. Therfore we  propose that    Bose-Einstein statistic  is the emergent statistics  both  in evolution of  haploid and diploid populations when  there is a  competition between processes enhancing genetic variation in the population (i.e. mutations or recombination process) and  selection  processes (which tend to decrease genetic variation in the population).

The genomic revolution started with the publication of the entire sequence  of the human genome \cite{Genome,Venter} has made possible the complete analysis of genome variations encoded in single nucleotide polymorphisms (SNPs). The complete set of SNPs of an individual characterizes together with the copy  number  variations  what  is unique about an individual. 
Variations in SNPs allelic states (i.e. different nucleotide composition of the SNPs) can affect how humans develop diseases and how they respond to pathogens or drugs.  It is well established \cite{Barabasi,Toroczkai,Sneppen,Bornholdt,Reka,Alon}, that genes are integrated in functional pathways and  interact through complex biological  networks.  Single nucleotide polymorphisms can  affect expression or function of the genes and they are correlated when gene products are part of a joint pathway.
If the functional pathways constitute the phenotype, then the interaction between the complete set of SNPs and these pathways encodes for the genotype-phenotype mapping. Consequently,  SNPs are interacting through the bio-molecular networks and their contribution to the fitness of an individual is  encoded in complex epistatic interactions between the SNPs \cite{Slatkin,Kishony}.

A recent paper has signed a turn-over in the study of epistatic interactions \cite{Landscape} . Indeed in \cite{Landscape} the epistatic network between a large set of pairs of mutations in  yeast,  has been fully characterized. This work demonstrates  the possibility of collecting data for  this   new fundamental type of  biological network \cite{Barabasi,Toroczkai,Sneppen,Bornholdt,Reka,Alon} in simple organisms,  by measuring the effect on fitness  of  pairs of mutations in yeast.
From the structural point of view, this network  is both modular \cite{Modular,Alon}  and fat tail \cite{Barabasi,Reka} as the regulatory network \cite{Alon}, the protein interaction network \cite{Protein,ProteinSK,Maritan} and the metabolic network \cite{Metabolic}.
From the evolutionary point of view, this  epistatic  network sheds light at the genotype-phenotype relation and  it  reveals a functional map of the cell in which genes with  highly correlated profiles delineate specific pathways.  
Similar networks exist also in higher organisms and, importantly, a substantial number of genes are regulated on the population level by the allelic states of polymorphic loci. A functional genome analysis of the signaling pathways of human thrombocytes revealed that a striking number of genes of the response cascade is under allelic control \cite{Blood}. 
 
\begin{figure}
\includegraphics[width=80mm, height=60mm]{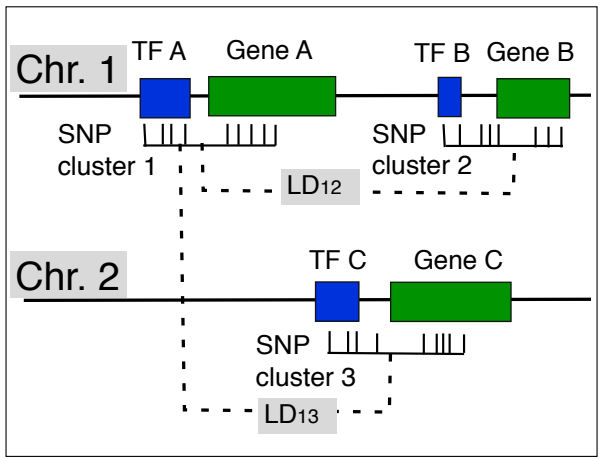}
\caption{(Color online) Linkage disequilibrium ($LD$) between clusters of SNPs.   
Single nucleotide polymorphisms are usually arranged in small clusters with all members being in complete linkage (haplotype). SNPs epistatic interactions are  mediated by  the transcription factors (TF) binding on regulatory regions and by the genes interacting in regulatory and signaling networks.   
Linkage disequilibrium between clusters of single nucleotide polymorphism (SNPs) can either appear  when two clusters are in close proximity  along a chromosome (as for example  $LD_{12}$ in the figure) or  when the SNPs clusters are at a significant distance along the chromosome or even  in two different chromosomes (as for example $LD_{13}$ in the figure).
While different  recombination rates might explain part of the linkage disequilibrium for clusters of SNPs in proximity along the chromosome, only epistatic interactions can explain  linkage disequilibrium for distant clusters. }
\label{SNPs}
\end{figure}

Linkage disequilibrium between SNPs \cite{Slatkin} is a key quantity for  identifying genes that are related to specific diseases.   In particular, linkage disequilibrium   indicates the non-random association of alleles at two or more loci (SNPs) and is widely observed through the human genome \cite{Lander}. 
Non random mating of a population and variation of the cross-over rate and finite evolutionary times contributes to the occurrence of  linkage disequilibrium in diploid populations.  However, also epistatic interactions between genetic loci  contributes to  the observed  linkage disequilibrium.
There is compelling evidence that linkage disequilibrium  occurs not only between genetic loci in close proximity to the chromosome but also between genetic loci at significant distance on the same chromosome or even on different chromosomes, as summarized in the scheme shown  in Fig. $\ref{SNPs}$. 
In order to explain this phenomenon it is necessary to describe the long-distance epistatic interactions between SNPs, which are not exclusively weak.

In order to develop an evolutionary theory in presence of epistatic interactions it is necessary to go  beyond the well defined single locus evolution \cite{Sigmund, Gillespie, Hartl,Sequence}. Nevertheless, most of the available mutiloci evolutionary theories  \cite{Hartl}  are typically  limited to a  few numbers of loci.  
A relevant exception is the recent paper of   R. Neher and  B. I. Shraiman \cite{Shraiman}  where the authors have  studied the role of the crossover rate  in an evolutionary theory of a large number of genetic loci, interacting  epistatically in a network.  Interestingly, they found by mean-field arguments and by numerical simulations, that the evolutionary model shows a  phase transition responsible for the level of genetic variation in a population. In fact, in their evolutionary theory, for high recombination rates the population is in the  "allele selection" phase in which genetic loci are only weakly associated,  while for low recombination rates the population  is in the "genotype selection" phase consisting of a set of competing genotypes locked in given allelic combinations.

In this paper we study the genetic variability  of sexually reproducing diploid populations where  free genetic recombination competes with  Darwinian natural selection \cite{Darwin}  under different    strengths of the selective pressure.  
We consider an epistatic network of $N$ loci (with $N\gg 1$) of a general topology, and we take the  fitness  of an individual as a  function of the allelic states of genetic loci in this epistatic network. In order to approach the formidable task to tackle the complexity of a mutiloci evolution, we neglect mutations (that  further contribute to the genetic variation in the population) and  we assume free recombination and infinite population and time limit. Finally, by making an ansatz on the shape of the distribution of gametes allelic state in the population, we are able to characterize all the stationary states in linkage equilibrium (while we leave to subsequent publications the study of solutions compatible with linkage disequilibrium).
The technical  improvement with respect to the previous  mutiloci theories that makes our theory exactly solvable, is due to the  advantages of defining the fitness function over an epistatic network and  using  the most recent developments of statistical mechanics
\cite{Bose,Parisi, Yedidia,Weigt, Mezard}.
In particular in this paper we make use of a self consistent argument \cite{Bose} combined with the cavity method \cite{Parisi,Yedidia,Weigt,Mezard}.

 The stationary states are multiple and therefore,  asymptotically in time, the state of a population depends on the initial conditions. 
Unexpectedly, the joint frequency of allelic states of pairs of linked loci, at stationarity, is  expressed in terms of a      Bose-Einstein distribution.
In a quantum Bose gas, the Bose-Einstein distribution describes the occupation number of energy levels. Moreover, a quantum Bose gas,  below a critical temperature, might undergo a Bose-Einstein condensation transition in which a finite fraction of particles are found in the ground state. 
In our mutiloci evolution dynamics for diploid populations,   the relation of the  steady state solution with the Bose-Einstein distribution, allows us to  predict a condensation transition in the   Bose-Einstein   universality class \cite{Huang,Quantumgas}.
 In this  evolutionary model, a pair of genetic loci is in the "ground state"  when they  are fixed,  i.e. when they are in a  given allelic state (not necessarily the same allelic state in each genetic locus) and they are not any more polymorphic. For  a given value of the  selective pressure, and a suitable topology of the epistatic network that allows for a  condensate phase, a finite fraction of pairs of loci is fixed. Therefore a finite fraction of pairs of loci is not any more polymorphic and the number of polymorphic loci is significantly reduced.  
 The basic mechanism behind the studied   condensation of genetic loci, is a  cooperative phenomenon in which, as the selection pressure increases,  one locus that is fixed in a beneficial configuration, affects the  other  linked loci inducing them to  fix in given allelic configurations, generating a macroscopic phase transition. A similar    phenomenon is also known in the two-loci evolutionary dynamics  and is called in the literature  "genetic hitchhiking" \cite{MS}.
 We note here that the phase transition observed in our model should be distinguished from  the "genotype selection" of \cite{Shraiman} because the genotypes in the population maintain a significant genetic variation due to the remaining polymorphic loci which have not been fixed. Interestingly, another phase transition between a polymorphic phase and a frozen phase has been numerically observed at a critical value of the mutation rate in sexual populations \cite{Peliti2}. 
\\
The paper is divided as follows: in Sec. II we  define the fitness function that  drives the  evolutionary dynamics in presence of a complex epistatic network; in Sec. III we define the evolutionary dynamic equation under consideration; in Sec. IV we  highlight the results regarding the steady state population of the evolutionary dynamics, including all the details of the  derivations in the subsequent appendixes;and  finally we give the conclusions.

\begin{figure}
\includegraphics[width=80mm, height=60mm]{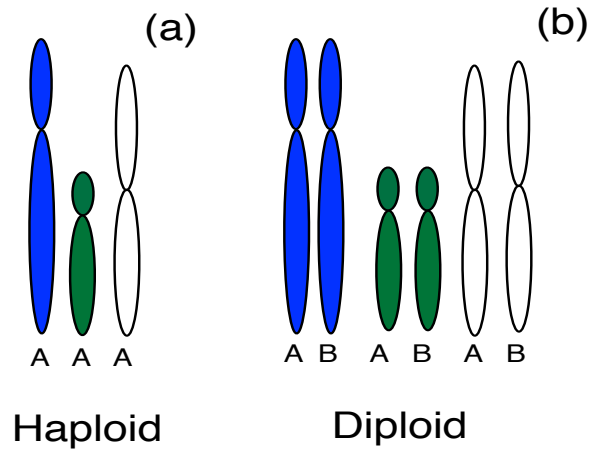}
\caption{(Color online) Difference between haploid and diploid cells: (a)  while haploid cells have   a single copy of each chromosome, diploid (b) organisms have two homologous copies   of each  chromosome.  }
\label{diploid.fig}
\end{figure}

\section{The fitness function and the   epistatic network }
 Haploid cells have a  single copy of each chromosome. On the contrary, diploid cells  have two homologous copies of each chromosome (see Fig. $\ref{diploid.fig}$).
Usually the  genome of each diploid individual  is given by the pairs of chromosomes $A$ and $B$ of  the two haploid   gametes coming  from the father and from the mother of the individual.
Let us suppose that   each gamete  is identified  by $N$ loci indicated with latin letters $i=1,2,\ldots, N$. If we indicate with $x_i$ the allelic state  at each locus  $i$,  the gamete is characterized by the sequence $\{x_i\}_{i=1,2,\ldots,N}$ with  $x_i=1,2,\ldots, Q$ and $Q$  given by the biochemistry of the DNA, i.e.  $Q=4$ indicating respectively the pair of ordered nucleotides AT (Adenine, Thymine), TA (thymine, adenine),  CG (cytosine, guanine) or GC (guanine, cytosine) in the double stranded DNA. 
Given this description of the gametes, each individual is characterized by the sequence $\{x_i^A,x_i^B\}_{i=1,2,\ldots,N}$ where  $x_i^{A/B}$ indicates the allelic states in each parental gametes  $A/B$.
In the  multiplicative non-epistatic (NE) scenario the fitness function $W^{NE}(\{x^A,x^B\})$  factorizes into  contributions coming from independent  single loci, i.e.  allelic states in a pair of loci do not have epistatic interactions.
In this traditional framework the fitness function is written as
\begin{equation}
W^{NE}(\{x^A,x^B\})=\prod_i \psi^{NE}_i(x_i^A,x_i^B).
\label{WNE}
\end{equation}
In the free recombination hypothesis, the minimal modification to this theory that is   compatible with the presence of epistatic interactions is that the fitness is a  function of allelic states of pairs of  loci.
Therefore we assume that the loci $i=1,2\ldots,N$ are linked in an epistatic network  formed by  $L$ links. We define by $\partial i$ the set of loci $j$ linked with locus $i$ in the network.
The epistatic couplings between pairs of loci  have a role in determining the fitness function that can be modified with respect to the single locus expression $(\ref{WNE})$, according to the expression
\begin{equation}
W(\{x^A,x^B\})=\prod_{<ij>}\phi_{ij}(x_i^A,x_j^A,x_i^B,x_j^B),
\label{W}
\end{equation}
where the product is extended to all  genetic loci $<i,j>$ linked in the epistatic network.
Therefore, the  fitness function in Eq. $(\ref{W})$ is the first non trivial correction to $(\ref{WNE})$ and includes a product of contributions coming from pairs of diploid allelic states at different loci.

We parametrize  the functions $\phi_{ij}(x_i^A,x_j^A,x_i^B,x_j^B)$ as in the following   
\begin{equation}
\phi_{ij}(x_i^A,x_j^A,x_i^B,x_j^B)=e^{-\beta U_{ij}(x_i^A,x_j^A,x_i^B,x_j^B)}
\label{U}
\end{equation}
where the parameter $\beta$ indicates the selective pressure. In fact  for $\beta=0$ we recover a neutral theory while for large $\beta$ small variations of the function $U_{ij}(x_i^A,x_j^A,x_i^B,x_j^B)$ yield large changes in the fitness.
Furthermore   the function $U_{ij}(x_i^A,x_j^A,x_i^B,x_j^B)$ has the following symmetries 
\begin{eqnarray}
U_{ij}(x_i,x_j,x_i^{\prime},x_j^{\prime})=U_{ji}(x_j,x_i,x_j^{\prime},x_i^{\prime})\nonumber \\
U_{ij}(x_i,x_j,x_i^{\prime},x_j^{\prime})=U_{ij}(x_i^{\prime},x_j^{\prime},x_i,x_j).
\label{pm}
\end{eqnarray}
with the last relation valid only if we assume perfect symmetry between the parental gametes, i.e. if we  exclude  to study the sex chromosomes $X,Y$.
The fitness landscape  defined in Eq. $(\ref{W})$ corresponds to a disordered Potts  Hamiltonian and therefore it  is in general  characterized by many local maxima.

\section{Evolutionary dynamics}

\begin{figure}
\includegraphics[width=80mm, height=100mm]{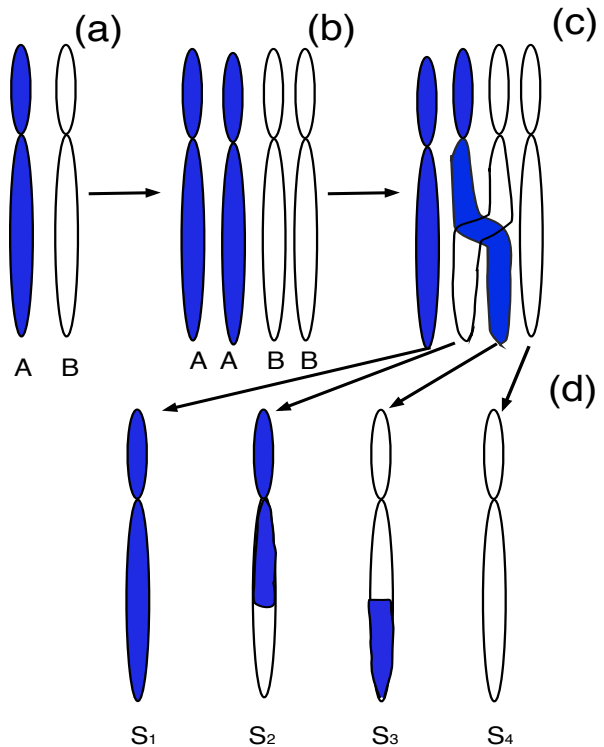}
\caption{(Color online) Graphical representation of meiosis.
Starting from a given diploid genome (a) of an individual AB, as the outcome of replication (b), cross-over and genetic recombination (c) a set of haploid gametes $S_1,S_2,\ldots S_n$ are formed (d). An exchange of genetic information might occur at each recombination site. In particular, every pair of genetic loci of the gamete either comse from a single parent or  from different parents.}
\label{recombination.fig}
\end{figure}

The evolutionary dynamics of diploid populations describes the information transfer of genetic information encoded in the gamete sequences.   
Each individual of a diploid population is carrying the information encoded in the gametes of their parents indicated as $A/B$.
The evolution is due to the transmission of each individual to the next generation of new gametes which are a random recombination of the information encoded in parental gametes $A/B$.
In physical terms the process can be seen as a "scattering" process in which two gametes $A/B$ generate a new individual ({\it fertilization}) and the new individual, if it reaches the reproductive state, carries  the information and gives rise (by {\it  meiosis}) to new gametes $S_1,S_2,\ldots S_n$ with $n=0,1,2\ldots$.
\begin{figure}
\includegraphics[width=80mm, height=60mm]{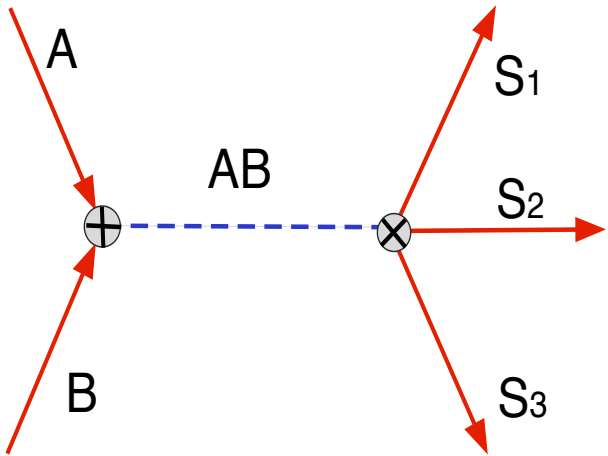}
\caption{(Color online) Graphical representation of the transfer of genetic information in overlapping generations  called in biology the {\it gametic life cycle}. This  diagram describes two parental  gametes $A, B$ (solid lines) that give rise to a zygote and then to an individual  $AB$  (dashed line) by a fertilization process indicated with  $\oplus$. This individual on his or her turn generates $n=3$ during meiosis $\otimes$ new gametes $ S_1,S_2,S_3$.}
\label{Feyman}
\end{figure}
We can visualize this process also called {\it gametic life cycle} by the diagram shown in  Fig. $\ref{Feyman}$ in which each solid line is a gamete and each dotted line is an individual.
The vertices of this diagram are indicated with a sign $\oplus$ when fertilization  occurs or with a sign $\otimes$ when meiosis occurs, i.e. a new gamete is generated by a process of recombination  of the diploid genetic information.
The presence of these vertices of the  diagram is an exclusive characteristics of diploid organisms wherever in haploid organism the single individual of the population is transmitting the genomic information to the next generation.  
The process of meiosis is a process of reduction in the genetic information of each diploid individual to generate gametes which have only half of the number of chromosomes.
During meiosis (see Fig. $\ref{recombination.fig}$) a  process of recombination can occur with small probability at given locations (recombination hotspots) on the chromosome. When a recombination event occurs, homologous sites on two chromosomes can mesh with one another and may exchange genetic information.
In our evolutionary dynamics we take the infinite population limit and we assume that at each genetic locus a recombination event can take place.
Therefore, the probability $P(\{x\})$ that a gamete has an allelic  configuration $\{x\}$, satisfies the  following dynamical equation
\begin{equation}
\hspace*{-5mm}\frac{\partial P(\{x\})}{\partial t}=M_{\{x\}}\left[\frac{W(\{x^A,x^B\})P(\{x^A\})P(\{x^B\})}{\Avg{W}}\right]-P(\{x\})
\label{me}
\end{equation}
where  $W(\{x^A,x^B\})$ is the fitness of the individual of diploid  allelic configuration $\{x^A,x^B\}$ given by Eqs. $(\ref{W})$ and $(\ref{U})$ and where $\Avg{W}$ is the average fitness
\begin{equation}
\avg{W}=\sum_{\{x^A\},\{x^B\}} W(\{x^A,x^B\})P(\{x^A\})P(\{x^B\}).
\label{Zp}
\end{equation}
The operator $M_{\{x\}}$ introduced in Eq. $(\ref{me})$ indicates the free recombination of genetic material occurring when a new gamete is generated.
In particular the operator  $M_{\{x\}}$  is defined as the average over the probability of free recombination   $\Pi(\{x\}|\{x^A,x^B\})$, i.e. the action of the operator over a generic function $f(\{x^A,x^B\})$ is given by
\begin{equation}
\hspace*{-5mm} M_{\{x\}}\left[f(\{x^A,x^B\})\right]=\sum_{\{x^A\},\{x^B\}}\Pi(\{x\}|\{x^A,x^B\})f(\{x^A,x^B\})
\end{equation}
with 
\begin{equation}
\Pi(\{x\}|{\{x^A,x^B\}})=\prod_{i=1}^N\left[\frac{1}{2}\delta(x_i,x_i^A)+\frac{1}{2}\delta(x_i,x_i^B)\right].
\end{equation}
We note here that in this model we assume 
 free recombination and 
equivalence between the parent gametes. Moreover,  in order to simplify the treatment of the evolutionary model, we limit  our study to evolution of diploid populations in absence of mutations.

\section{Results}

\subsection{General form of steady state  probability distributions}

If the network is locally tree-like,  the general  structure of the solution to the evolutionary equation $(\ref{me})$ is given by  
\begin{equation}
{P}(\{x\})=\sum_h \pi(h) \prod_{<i,j>}b_{ij}^h(x_i,x_j).
\label{Gtree}
\end{equation}
where $<i,j>$ indicates all the pairs of linked nodes  present in the epistatic network.

In our model the fitness function of the type $(\ref{W})$ and  $(\ref{U})$, for any given epistatic network, is static and bounded from above, therefore we always expect to find asymptotically in time the population in a stationary state given by the solution of  to the equation
\begin{equation}
P(\{x\})=M_{\{x\}}\left[\frac{W(\{x^A,x^B\})P(\{x^A\})P(\{x^B\})}{\Avg{W}}\right].
\label{mes}
\end{equation}
These stationary states, do not necessarily correspond to a maximum of the fitness \cite{Moran}.
In the case treated in this paper,  in which  the epistatic network is fixed and   locally  tree-like, we can find, for every generic fitness function of  type $(\ref{W})$ and $(\ref{U})$, the possible stationary states of the population (see appendixes) of the type
\begin{equation}
P(\{x\})=\prod_{<i,j>}b_{ij}(x_i,x_j),
\label{Pb}
\end{equation}
where the product is extended to all  genetic loci $<i,j>$ linked in the epistatic network.
The type of solutions $(\ref{Pb})$ is a subset of the general type of solutions $(\ref{Gtree})$.
In particular, in order for $(\ref{Pb})$ to be a solution of the stationary relation Eq. $(\ref{mes})$, $P(\{x\})$ must satisfy the condition 
\begin{equation}
P(\{x^A\})P(\{x^B\})=P(\{x^A_{-i}\},x_i^B) P(\{x^B_{-i}\},x_i^A)
\end{equation}
where $\{x^{A}_{-i}\}$ indicates all the variables $\{x^A\}$ except variable $\{x_i^A\}$ and $\{x^{B}_{-i}\}$ indicates all the variables $\{x^B\}$ except variable $\{x_i^B\}$.
It can be easily shown that these conditions, enforce  linkage equilibrium between allelic states.
In this paper we restrict our attention to this type of solutions and we  leave to subsequent publications the study of stationary state distributions compatible with linkage disequilibrium.

The marginal frequencies $p_{ij}(x_i,x_j)$ of allelic states on   pairs of linked loci $(i,j)$ are defined as 
\begin{equation}
p_{ij} (x_i^{\prime},x_j^{\prime}) =\sum_{\{x\}} P(\{x\})\delta(x_i,x_i^{\prime}) \delta(x_j,x_j^{\prime}).
\label{marginal}
\end{equation}
\subsection{Bose-Einstein distribution}
In order to find all the stationary solutions solving Eq. $(\ref{mes})$ of the form given by $(\ref{Pb})$, we used  a self-consistent argument \cite{Bose} combined with the cavity method \cite{Parisi,Yedidia,Mezard}.
In particular  we find that in the stationary state, not necessary a maximum of the fitness function \cite{Moran}, the marginal frequencies $p_{ij}(x_i,x_j)$  defined in Eq. $(\ref{marginal})$ are given by (see Appendix $\ref{C} and \ref{D}$) 
\begin{equation}
p_{ij}(x_i,x_j)=\frac{1}{Z}G_{ij}(x_i,x_j)\left\{1+n_B[\epsilon_{ij}(x_i,x_j)]\right\}
\label{fre}
\end{equation}
if  $\epsilon_{ij}(x_i,x_j)>\mu$. The functions  $\epsilon_{ij}(x_i,x_j), G_{ij}(x_i,x_j)$ in Eq. $(\ref{fre})$ and the constants $\mu,Z$ can  be derived from the self consistent solution of the stationary state of the evolutionary dynamics described by Eq. $(\ref{me})$  (see Appendix $\ref{C}$).
In Eq. $(\ref{fre})$ $n_B(\epsilon)$ indicates the Bose-Einstein occupation number and $\mu$ indicates the "chemical potential" of the evolutionary dynamics. The Bose-Einstein occupation number is  defined  as
\begin{eqnarray}
n_B[\epsilon_{ij}(x_i,x_j)]&=&\frac{1}{e^{\beta[\epsilon_{ij}(x_i,x_j)-\mu]}-1},\label{Bose}
\end{eqnarray} Equation  $(\ref{fre})$  relates the joint probability of pair of linked loci with a   Bose-Einstein distribution arising in the study of  quantum Bose gases \cite{Huang,Quantumgas}.
Here the functions, $\epsilon_{ij}(x_i,x_j)$ play the role of "energy states" of this Bose-Einstein distribution. These functions are not known a priori but they are the outcome of the evolutionary dynamics. 
A relevant aspect of this solution is that we might find several different sets of functions $\{\epsilon_{ij}(x_i,x_j),G_{ij}(x_i,x_j)\}$ and variables $Z,\mu$ that satisfy the stationary condition of the evolutionary dynamics.
These different solutions have to be identified with different possible populations of a given species. In fact, given different initial conditions the population evolving according to Eq. $(\ref{me})$ can be found, asymptotically in time,  in different stationary states.  According to the general mutiloci evolutionary scenario \cite{Moran}, these  steady states   do not in general correspond to  local maxima of the fitness landscape.\\

 Interestingly,  the marginal frequencies $p_{ij}(x_i,x_j)$ are significantly modified if  a given pair of  allelic configuration $(x_i^{\star},x_j^{\star})$ reaches the minimal allowed  "energy level" $\epsilon_{ij}(x_i^{\star},x_j^{\star})=\mu$. In this case we found $G_{ij}(x_i,x_j)=0$ for every allelic state $(x_i,x_j)$ and the pairs of linked loci $(i,j)$  gets fixed (see Appendix $\ref{E}$).
Therefore if  $\epsilon_{ij}(x_i^{\star},x_j^{\star})=\mu$, the joint probability $p_{ij}(x_i,x_j)$ is given by 
\begin{equation}
p_{ij}(x_i,x_j)=\delta(x_i,x_i^{\star})\delta(x_j,x_j^{\star}).
\label{fix}
\end{equation} 
 To be specific in the terminology used, here and in the following  we assume that a  pair of genetic loci is fixed if and only the joint distribution is given by $(\ref{fix})$, i.e.  if and only if both genetic loci ($i$ and $j$) are fixed.

\begin{figure}
\includegraphics[width=80mm, height=60mm]{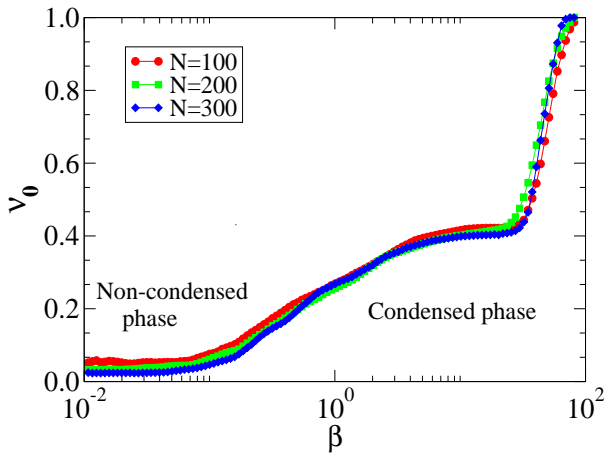}
\caption{(Color online) Numerical evidence for the condensation phase transition. The fraction  $\nu_0$ of fixed pairs of loci is plotted as a function of the selective pressure $\beta$. The data are averaged over $50$ random fitness realizations and are shown for a epistatic network  with degree distribution given by Eq. $(\ref{p})$.
The   fitness function  used is given by  Eqs. $(\ref{W})$ and $(\ref{U})$ with the matrix elements $U_{ij}(x_i^A,x_j^A,x_i^B,x_j^B)$  satisfying the symmetry constraints $(\ref{pm})$ and drawn randomly from a uniform distribution in the interval $(0,1)$. The data are shown for a  number $N$ of genetic loci with   $N=100,200,300$.}
\label{Transition}
\end{figure}

\subsection{Condensation transition}

 The joint distributions $(\ref{fre})$ and $(\ref{fix})$ must always satisfy the normalization constraints
\begin{equation}
1=\sum_{x_i,x_j}p_{ij}(x_i,x_j)
\label{uno}
\end{equation}
valid for every pair of linked loci $(i,j)$.
This set of equations plays the role  of "conservation" laws for the evolutionary dynamics and  determines the phase diagram of the evolutionary dynamics.
Inserting in $(\ref{uno})$ the expression $(\ref{fre})$ for $p_{ij}(x_i,x_j)$ when $\epsilon_{ij}(x_i,x_j)<\mu$ and expression  $(\ref{fix})$ for $p_{ij}(x_i,x_j)$ when $\epsilon_{ij}(x_i,x_j)=\mu$
we can write the normalization conditions $(\ref{uno})$ as in the following,
\begin{eqnarray}
\label{self}
(1-\nu_{ij})&=&\frac{1}{Z}\sum_{x_i,x_j} G_{ij}(x_i,x_j)+ \\
&&\frac{1}{Z}\sum_{\begin{array}{c}x_i,x_j \\\epsilon_{ij}(x_i,x_j)>\mu \end{array}}G_{ij}(x_i,x_j)n_B[\epsilon_{ij}(x_i,x_j)]\nonumber
\end{eqnarray}
valid for every pair of linked loci $(i,j)$. In Eq. $(\ref{self})$,  $n_B(\epsilon)$ is the Bose-Einstein distribution defined in Eq. $(\ref{Bose})$ and the matrix elements $\nu_{ij}$ are  $\nu_{ij}=1$  the pairs of linked loci $(i,j)$ are fixed, otherwise $\nu_{ij}=0$. 
Considering Eq.  $(\ref{self})$ and   averaging over all pairs of linked loci we can write 
\begin{equation}
(1-\nu_0)=\frac{1}{Z}\int_{\epsilon>\mu} d\epsilon [1+n_{B}(\epsilon)]g_{\beta}(\epsilon)
\label{selfc}
\end{equation}
where  $n_B(\epsilon)$ is the Bose-Einstein distribution defined in Eq. $(\ref{Bose})$,  and $\nu_0$ is the fraction of pairs of linked loci that are  fixed in the population. The quantities $g_{\beta}(\epsilon)$ and $\nu_0$ present in Eq. $(\ref{selfc})$ are  given by 
\begin{eqnarray}
\hspace*{-5mm}g_{\beta}(\epsilon)&=&\lim_{\Delta \epsilon \rightarrow 0}\frac{1}{\Delta\epsilon}\frac{1}{2L}\sum_{i,j\in \partial i}\sum_{x_i,x_j}G_{ij}(x_i,x_j)\chi_{\Delta \epsilon}(\epsilon_{ij}(x_i,x_j)-\epsilon)\nonumber\\
\nu_0&=&\frac{1}{2L}\sum_{i,j\in \partial i}\sum_{x_i,x_j}\chi_0[\epsilon_{ij}(x_i,x_j)-\mu_c].
\label{gbetae}
\end{eqnarray}
where $\chi_\delta(x)=1$ if $|x|\leq \delta$ and $\chi_{\delta}(x)=0$ otherwise. 
Depending on the form of $g_{\beta}(\epsilon)$ the solution of Eq. $(\ref{selfc})$ might indicate the occurrence, as a function of $\beta$, of a  condensation phase transition  characterized by the order parameter $\nu_0$.  For epistatic network topologies and fitness functions which display this phase transition,  we can distinguish, as a function of $\beta$, between a noncondensed phase  in which the fraction of fixed pairs of loci $\nu_0$ vanishes in the thermodynamic limit, i.e.  $\nu_0\rightarrow 0$ as $N\to \infty$ and a "condensed phase" in which the fraction of fixed pairs of genetic loci is finite in the thermodynamic limit, i.e.  $\nu_0 \rightarrow \overline{\nu}>0$ as $N\to \infty$. 

This condensation phase transition  is in the same universality class of the Bose-Einstein condensation transition as it depends on the value of the integral of a Bose-Einstein distribution present in  Eq. $(\ref{selfc})$.
We observe, nevertheless, that Eq. $(\ref{selfc})$ differs from the equation fixing the average number of particles in a Bose gas \cite{Huang,Quantumgas} because  the function $g_{\beta}(\epsilon)$ given by Eq. $(\ref{gbetae})$  depends on $\beta$ while the correspondent density of states  in a quantum Bose gas is independent of $\beta$. Moreover in Eq. $(\ref{gbetae})$ there is an additional factor $Z$ in the right hand  side with respect to the correspondent equation in the quantum Bose gas. 

In the noncondensed phase all genetic loci are polymorphic, on the contrary, in the "condensed phase" only a fraction of genetic loci is polymorphic. In which phase are  diploid populations usually found?
If we assume that each base of the DNA is a candidate SNP, we observe that polymorphisms only occur in a finite fraction of bases. For example, in the human genome less than 1\% of the bases corresponds to SNPs.  Here we propose   that the condensation of genetic loci due to epistatic interactions, might  significantly contribute to   the reduction in genetic variation  within a species. 

\subsection{Numerical evidence of the condensation transition}
While the results exposed in the preceding section are valid for any fitness function of  type $(\ref{W})$ and  tree-like epistatic network, the actual phase diagram of the evolutionary dynamics might change depending on the topology of the network and on the detail of the fitness function. In this paragraph we show numerical evidence for the   condensation of genetic loci by solving the self-consistent equations (see Appendix $\ref{C}$) that determine  $\{\epsilon_{ij}(x_i,x_j),G_{ij}(x_i,x_j)\}$ and  $\nu_0$, for a given fitness function, starting from random initial conditions. In particular we consider a network topology  that  allows for long-distance epistatic interactions (see Fig. $\ref{SNPs}$). We have therefore chosen to study  an epistatic     network  with degree distribution 
\begin{equation}
P(k)\propto k^{-\gamma}.
\label{p}
\end{equation}  In Fig. $\ref{Transition}$ we  show evidence for the occurrence of the condensation transition of genetic loci when the epistatic network is a random network with degree distribution given by $(\ref{p})$ and  $\gamma=3$. 
In particular in Fig. $\ref{Transition}$ we have plotted  the  fraction $\nu_0$ of fixed pairs of loci (averaged over several random realizations of the fitness function) as a function of the evolutionary pressure $\beta$.

 The "condensed phase" is defined as the region  where $\nu_0$ is large and does not show finite size effects. Outside this region, instead, we have the "non condensed phase" where the fraction of fixed loci goes to zero in the limit of large $N$, i.e. $\nu_0\rightarrow 0$ as $N\rightarrow \infty$. 
The   fitness function  used in the numerical solution reported in Fig. $\ref{Transition}$,  is given by  Eqs. $(\ref{W})$ and $(\ref{U})$ with the matrix elements $U_{ij}(x_i^A,x_j^A,x_i^B,x_j^B)$  satisfying the symmetry constraints $(\ref{pm})$ and drawn randomly from a uniform distribution in the interval $(0,1)$.  Finally, in order to reduce the  time for the numerical solution of the self-consistent equations we have taken $Q=2$.

  \subsection{Condensation transitions in evolutionary dynamics}
  
Condensation phase transitions universally occur in evolutionary models.
A pivotal condensation phase transition occurs in the  quasispecies \cite{Eigen,Peliti,Nowak} evolutionary model of haploid populations that describes the competition between random mutations, which tend to increase the genetic variation of the population, and natural selection  which tends to reduce it.
In the quasispecies model, when the mutation rate $\mu$ is less than a critical value $\mu_c$, i.e. $\mu<\mu_c$, the haploid population is localized in the sequence space, and when, instead, $\mu>\mu_c$ there is no possibility to define a typical sequence in the population.
A condensation transition also occurs in the  "house of cards" model of Kingman \cite{Kingman} which describes the quasispecies model in the limit of infinite loci.
Kingman   characterized  the condensation transition in the "house of cards" model but only recently, with the study of evolving complex systems, i.e networks \cite{Bose} and ecosystems \cite{Ferretti}, and in a more elaborated model with pleiotropy \cite{Kadanoff}  it was recognized that this condensation can be mapped to a Bose-Einstein condensation in a Bose gas. 

Condensation phase transitions also occur in diploid populations. In \cite{Shraiman} it was shown that the phase transition between the "allele selection" phase and the "genotype selection" phase is a condensation phase transition below which, for low recombination rates, few genotypes are selected in the population.

Here we show that a condensation transition in the Bose-Einstein universality class is also occurring in diploid populations in  the  presence of  free genetic recombination.
The  novelty of this transition is that  the  occurrence of the Bose-Einstein statistics is not caused by mutations (as it is the case for the quasispecies and the "house of cards" models) but only by genetic recombination. Moreover, this condensation transition differs from the transition between "allelic selection" and "genotype selection"  of \cite{Shraiman} because  in the condensed phase of the present  mutiloci evolutionary theory, the population maintains a wide variation although the number of polymorphic loci is significantly reduced.
Finally, the condensation of genetic loci of diploid populations  is a consequence of the non-trivial interactions of genetic loci in the epistatic network while in the quasispecies model and in the "house of cards" model the interactions between the individuals of the population are only mediated by the competition for finite resources.
Therefore the condensation of genetic loci in the present evolutionary theory  relates to the condensation transition in the quasispecies model \cite{Eigen,Nowak} as the condensation transition in interacting quantum Bose gases \cite{Quantumgas} relates to the condensation transition in non-interacting quantum Bose gases \cite{Huang,Quantumgas}.
Finally, it is   fascinating  to observe how different are the  underlying  mechanisms yielding to condensation transitions in haploid and diploid populations while both mechanisms have been  selected by nature for their  evolutionary advantages.

\section{Conclusions}
In conclusion we have studied  a mutiloci evolutionary dynamics in  sexually reproducing diploid populations in which random genetic recombination tends to increase genetic variation while natural selection tends to reduce it. The  mutiloci evolution  is driven by a fitness function  defined on an epistatic  network of genetic loci.
We have found that the  stationary states of this evolutionary dynamics are multiple, and depend on the initial condition of the population.
Unexpectedly, we have found   that the joint distributions of allelic states at linked loci, can, at stationary state,  be expressed in terms of a Bose-Einstein  distribution with the "energy levels" depending on the network of epistatic interactions between genetic loci.
The relation of the joint distributions with the Bose-Einstein distribution allows us to define a possible condensation phase transition in the universality class of the Bose-Einstein condensation. 
Below this condensation phase transition a finite fraction of pairs of genetic loci  is fixed in the population and the number of polymorphic loci is strongly reduced.
Therefore  we propose  here the Bose-Einstein condensation of genetic loci as a possible mechanism contributing to  the reduction in genetic variation  within a species.

In the future it is promising to include in this model the role of mutations (that increase genetic variation in the population), finite populations  (that contribute to the existence of linkage disequilibrium)  and the adaptive nature of the epistatic network.
Moreover, we plan in future works  to   include in the model the possibility for a variable crossover rate  and to go beyond the assumption of a locally tree-like epistatic network.  Finally it would be interesting  to characterize further the relation between the evolutionary dynamics and quantum mechanics \cite{Davies,Lloyd} by investigating the role of condensation transitions present in evolution \cite{Eigen,Peliti,Nowak,Kingman,Bose,Kadanoff,Ferretti,Shraiman} belonging to  the Bose-Einstein universality class.

\appendix

\section{Calculations of marginals using the cavity method (Bethe-Peierls approximation)}
\label{B}

\begin{figure}
\includegraphics[width=80mm, height=60mm]{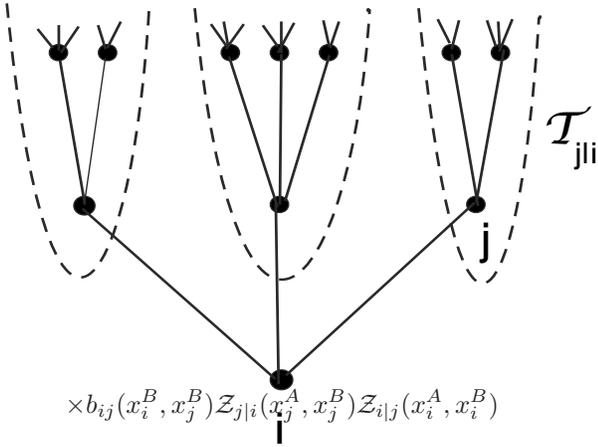}
\caption{Iteration tree for the calculation of the partition function. The sub-trees rooted in all $j \in \partial i$ are marked with dashed lines}
\label{tree_fig}
\end{figure}

\begin{figure}
\includegraphics[width=80mm, height=60mm]{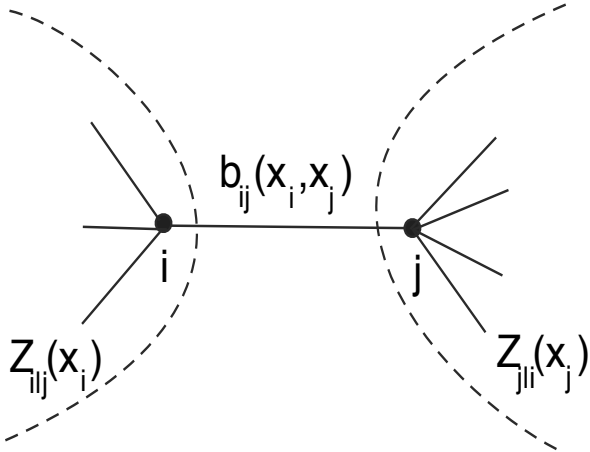}
\caption{The calculation of the marginal distribution $p_{ij}(x_i,x_j)$ can be expressed in terms on the restricted partition functions of the trees ${\cal T}_{i|j}$  and ${\cal T}_{i|j}$ according to Eq. $(\ref{Am})$.}
\label{tree2_fig}
\end{figure}

In the  hypothesis that the epistatic network is  locally tree-like 
we look for  solution to the evolutionary equation $(\ref{me})$ of the form given by 
\begin{equation}
P(\{x\})=\prod_{<i,j>}b_{ij}(x_i,x_j)
\label{Ap1}
\end{equation}
with $b_{ij}(x_i,x_j)$ to be determined by  Eq. $(\ref{me})$.
The marginal  frequency   $p_{ij}(x_i,x_j)$ of a pair of linked  loci is defined as 
\begin{equation}
p_{ij}(x_i^{\prime},x_j^{\prime})=\sum_{\{x\}} P(\{x\}) \delta(x_i,x_i^{\prime}) \delta(x_j,x_j^{\prime}).
\label{Ap} 
\end{equation}
If we assume to know the functions $b_{ij}(x_i,x_j)$ the marginal frequencies $(\ref{Ap})$ can be calculated by the cavity method \cite{Yedidia,Mezard,Parisi,Weigt} (or the Bethe Peierls approximation) exactly valid on locally tree-like networks.
For completeness we describe here the fundamentals of the  cavity method    that will be used in the following derivation of the stationary state solution of the evolutionary dynamics defined in $(\ref{me})$.
Let us work in the hypothesis that the network is really a tree.
Consider a known distribution function of the type $(\ref{Ap1})$ with given functions $b_{ij}(x_i,x_j)$. Our purpose  is to calculate in an efficient way  the marginals  $p_{ij}(x_i,x_j)$ defined in  $(\ref{marginal})$.
These distributions can be calculated by a simple iterative procedure.
Let us introduce the restricted partition functions $Z_{j|i}(x_j)$ of the sub-trees ${\cal T}_{j|i}$ rooted in the node  $j\in \partial i$ and not including node $i$. These subtrees are marked by dashed lines in Fig. $\ref{tree_fig}$.  The restricted partition functions $Z_{j|i}(x_j)$ is defined as
\begin{equation}
Z_{j|i}(x_j)=\sum_{\begin{array} {c}\{x_{\ell}\}\\
\ell\in {\cal T}_{j|i}\setminus j\end{array}}\prod_{\ell,\ell^{\prime}\in {\cal T}_{j|i}}b_{\ell, \ell^{\prime}}(x_{\ell},x_{{\ell}^{\prime}}),
\end{equation}
where the sum is performed over all the variables $x_{\ell}$ associated with the nodes $\ell$ of the sub-tree ${\cal T}_{j|i}$ except for the variable $x_j$.
 Using this definition and the assumption that the network is locally a tree (see Fig. $\ref{tree2_fig}$), it  is easy to prove that the marginal distributions  $p_{ij}(x_i,x_j)$ defined in $(\ref{marginal})$ are given by
\begin{equation}
p_{ij}(x_i,x_j)=b_{ij}(x_i,x_j)Z_{j|i}(x_j) Z_{i|j}(x_i).
\label{Am}
\end{equation}
In order to calculate the restricted partition functions $Z_{j|i}(x_j)$ we use the following recursive equation, that expresses the relation between restricted partition functions  of nested subtrees, 
\begin{equation}
Z_{j|i}(x_j)=\prod_{k\in \partial j\setminus i}\sum_{x_k }b(x_j,x_k) Z_{k|j}(x_k).
\label{Arec}
\end{equation}
These recursive equations are sufficient to define the full set of restricted partition functions ${\cal Z}_{j|i}(x_j)$ within a constant that must be fixed by the normalization conditions
\begin{equation}
\sum_{x_i,x_j}p_{ij}(x_i,x_j)=1.
\label{ANN}
\end{equation}

The cavity method is proved to be exact  not only on trees but also on locally tree-like networks.
Nevertheless it also generally used for networks with short loops as long as the recursive equations $(\ref{Arec})$ have a solution.
We can extend this formalism to generic distributions defined on locally tree-like networks in which each node is associated to more than one variable.
Let us for example consider the case of the distribution  function ${\cal P}(\{x^A,x^B\})$ defined in terms of the distribution $P(\{x\})$ and is given by 
\begin{equation}
{\cal P}(\{x^A,x^B\})=\left[\frac{W(\{x^A,x^B\})P(\{x^A\})P(\{x^B\})}{\Avg{W}}\right].
\label{APcal}
\end{equation}
Also this distribution function, like the distribution $P(\{x\})$, is  defined on a tree,  but in this case to  each node $i$, where two variables are associated: $x_i^A$ and $x_i^B$ .
Assuming that the distribution $P(\{x\})$ is given by Eq. $(\ref{Ap1})$, when the functions $b_{ij}(x_i,x_j)$  are known,  we can write the distribution ${\cal P}(\{x^A,x^B\})$ as a product of terms depending on indeces $(i,j)$ of linked pair of nodes according to the expression,
\begin{eqnarray}
{\cal P}(\{x^A,x^B\})&=&\prod_{<i,j>}\frac{e^{-\beta U_{ij}(x_i^A,x_j^A,x_i^B,x_j^B)}}{\Avg{W}}
\nonumber \\
&&\times \prod_{<i,j>} b_{ij}(x_i^A,x_j^A)b_{ij}(x_i^B,x_j^B).
\label{APcal2}
\end{eqnarray}
 Therefore, proceeding as in the previous case, we can use the cavity method and  define  the restricted partition functions ${\cal Z}_{j|i}(x_j^A,x_j^B)$ defined on the subtrees  ${\cal T}_{j|i}$ and determined within a constant by  the recursive equations
\begin{eqnarray}
{\cal Z}_{j|i}(x_j^A,x_j^B)&=&\prod_{k\in \partial j\setminus i}\sum_{x^A_k,x^B_k}e^{-\beta U_{ik}(x_j^A,x_k^A,x_j^B,x_k^B)}\nonumber\\
&&\hspace*{-12mm}\times b_{ik}(x_j^A,x_k^A)b_{ik}(x_j^B,x_k^B){\cal Z}_{k|j}(x_k^A,x_k^B).
\label{Aca}
\end{eqnarray} 
Finally the marginals $m_{ij}(x_i,x_j,x_i^{\prime},x_j^{\prime})$ of the probability distribution ${\cal P}(\{x^A,x^B\})$ are defined as
\begin{eqnarray}
m_{ij}(x_i,x_j,x_i^{\prime},x_j^{\prime})&=&\sum_{\{x^A\},\{x^B\}}{\cal P}(\{x^A,x^B\})\delta(x_i^A,x_i)\nonumber \\ & & \hspace*{-8mm}\times\delta(x_i^B,x_i^{\prime})\delta(x_j^A,x_j)\delta(x_j^B,x_j^{\prime})
\end{eqnarray}
and are given in terms of the restricted partition functions according to the following relation
\begin{eqnarray}
m_{ij}(x_i^A,x_i^B,x_j^A,x_j^B)&=&\frac{1}{Z}e^{-\beta U_{ij}(x_i^A,x_j^A,x_i^B,x_j^B)}b_{ij}(x_i^A,x_j^A)\nonumber \\
&&\hspace*{-119mm}\times b_{ij}(x_i^B,x_j^B) {\cal Z}_{j|i}(x_j^A,x^B_j){\cal Z}_{i|j}(x_i^A,x^B_i)
\label{AMb}
\end{eqnarray}
where the normalization constant $Z$ can be calculated by imposing the normalization conditions
\begin{equation}
\sum_{x_i^A,x_j^A,x_i^{B},x_j^{B}}m_{ij}(x_i^A,x_j^A,x_i^B,x_j^B)=1. 
\label{ANM}
\end{equation}

 \section{Characterization  of the steady state solution} 
 \label{C}
 The stationary states of Eq. $(\ref{me})$ are  given by the solutions to the equation
\begin{equation}
P(\{x\})=M_{\{x\}}\left[\frac{W(\{x^A,x^B\})P(\{x^A\})P(\{x^B\})}{\Avg{W}}\right].
\label{Ames}
\end{equation}
If the network of  epistatic interactions is  locally  tree-like,  we can find the exact   solution of Eq. $(\ref{Ames})$ using  a self-consistent argument \cite{Bose} combined with the cavity method   \cite{Yedidia,Mezard,Parisi,Weigt}.

In our self-consistent assumption we suppose to know the functions $b_{ij}(x_i,x_j)$ determining the distribution $P(\{x\})$ given by Eq. $(\ref{Ap1})$ and the distribution ${\cal P}(x^A,x^B)$ defined in Eqs. $(\ref{APcal})$ and  $(\ref{APcal2})$,  both present in Eq. $(\ref{Ames})$.
If we suppose to know the functions $b_{ij}(x_i,x_j)$ we can evaluate the marginal distributions $p_{ij}(x_i,x_j)$ and  $m_{ij}(x_i,x_j,x_i^{\prime},x_j^{\prime})$ by the cavity method as described in the previous section.
Finally imposing the stability condition $(\ref{Ames})$, on the marginal distributions, taking into account for the free recombination operator $M_{\{x\}}$ we get 
\begin{eqnarray}
b_{ij}(x_i,x_j)Z_{i|j}(x_i)Z_{j|i}(x_j)&=&\frac{1}{2}\sum_{x_i^B,x_j^B}m_{ij}(x_i,x_j,x_i^B,x_j^B)\nonumber \\
&&\hspace*{-12mm}+\frac{1}{2}\sum_{x_i^B,x_j^A}m_{ij}(x_i,x_j^A,x_i^B,x_j)
\label{Aself0}
\end{eqnarray}
The first   term in the right hand side of Eq. $(\ref{Aself0})$ comes from the probability that both allelic states $i$ and $j$ derive from a single parent. The second  term in the right hand side of Eq. $(\ref{Aself0})$, instead, describes the  probability of a cross-over of genetic information, i.e. the event that the two allelic states $(i,j)$ of the new gamete originate  from different parental gametes. In Eq. $(\ref{Aself0})$ we have used the fact that  the marginals $m_{ij}(x_i,x_j,x_i^{\prime},x_j^{\prime})$ are symmetric, i.e.
\begin{equation}
m_{ij}(x_i,x_j,x_i^{\prime},x_j^{\prime})=m_{ji}(x_i^{\prime},x_j^{\prime},x_i,x_j)
\end{equation}
as a consequence of the assumed symmetries of the fitness function given by  $(\ref{pm})$.
Using Eq. $(\ref{AMb})$ to express the marginals, explicitly taking into account of the dependence of the right hand side of Eq. $(\ref{Aself0})$ on  $b_{ij}(x_i,x_j)$, we can write Eq. $(\ref{Aself0})$ as in the following 
\begin{eqnarray}
Zb_{ij}(x_i,x_j)Z_{i|j}(x_i)Z_{j|i}(x_j)&=&b_{ij}(x_i,x_j){F_{ij}(x_i,x_j)}\nonumber \\
&&+{G_{ij}(x_i,x_j)}.
\label{Aself}
\end{eqnarray}
where  the functions $F_{ij}(x_i,x_j),G_{ij}(x_i,x_j)$ are defined as 
\begin{eqnarray}
F_{ij}(x_i,x_j)&=&\frac{1}{2}\sum_{x^B_i,x^B_j}\frac{1}{b_{ij}(x_i,x_j)} m_{ij}(x_i,x_j,x_i^B,x_j^B)\nonumber\\
&&\hspace*{-15mm}+\frac{1}{2}\frac{1}{b_{ij}(x_i,x_j)}m_{ij}(x_i,x_j,x_i,x_j), \nonumber \\
G_{ij}(x_i,x_j)&=&\frac{1}{2}\sum_{x^A_j,x^B_i}m_{ij}(x_i,x_j^A,x_i^B,x_j)\nonumber\\
& & \times\left[1-\delta(x_i,x_i^B)\delta(x_j^A,x_j)\right].
\label{Adef}
\end{eqnarray}
These functions  can be calculated by the cavity method in terms of the restricted partition functions ${\cal Z}_{i|j}(x_i)$ satisfying Eq.$(\ref{Aca})$ as expressed by the following equations,
\begin{eqnarray}
F_{ij}(x_i,x_j)&=&\frac{1}{2}\sum_{x^B_i,x^B_j}e^{-\beta U_{ij}(x_i,x_j,x_i^B,x_j^B)}b_{ij}(x_i^B,x_j^B) \nonumber \\
&&\times {\cal Z}_{j|i}(x_j,x^B_j){\cal Z}_{i|j}(x_i,x^B_i)+\nonumber \\
&&+\frac{1}{2}\,e^{-\beta U_{ij}(x_i,x_j,x_i,x_j)}b_{ij}(x_i,x_j)\nonumber \\
&&\times {\cal Z}_{j|i}(x_j,x_j) {\cal Z}_{i|j}(x_i,x_i)\nonumber \\
G_{ij}(x_i,x_j)&=&\frac{1}{2}\sum_{x^A_j,x^B_i}e^{-\beta U_{ij}(x_i,x_j^A,x_i^B,x_j)}b_{ij}(x_i,x_j^A)\nonumber \\
&&\times b_{ij}(x_i^B,x_j){\cal Z}_{j|i}(x_j^A,x_j) {\cal Z}_{i|j}(x_i,x^B_i)\times\nonumber\\
& & \times\left[1-\delta(x_i,x_i^B)\delta(x_j^A,x_j)\right],\nonumber \\
&&\hspace*{-20mm} Z=\sum_{x_i,x_j}b_{ij}(x_i,x_j)F_{ij}(x_i,x_j)+G_{ij}(x_i,x_j).
\label{Adef3}
\end{eqnarray}
The Equations $(\ref{Aself})$, can be seen as a set of equations able to determine self-consistently the functions $b_{ij}(x_i,x_j)$ closing the self-consistent argument.
The coupled equations $(\ref{Arec}),(\ref{ANN}),(\ref{Aca}),(\ref{Aself})$ and $(\ref{Adef3})$ provide the solution for the stationary state of the mutiloci evolution.
These cavity equations will in general  lead to multiple solutions corresponding to the multiplicity of possible steady states of the studied  evolutionary dynamics.

\section{Bose-Einstein distribution}
\label{D}
We want here to comment on the structure of the stationary distribution  found by the solution of the mutiloci evolution provided in the previous section.
Solving Eqs. $(\ref{Aself})$ for $b_{ij}(x_i,x_j)$ yields
\begin{equation}
b_{ij}(x_i,x_j)=\frac{G_{ij}(x_i,x_j)}{{Z}Z_{i|j}(x_i)Z_{j|i}(x_j)-F_{ij}(x_i,x_j)},
\label{A16}
\end{equation}
as long as  $\left[{Z}Z_{i|j}(x_i)Z_{j|i}(x_j)-F_{ij}(x_i,x_j)\right]>0$. Let us for the moment assume that this last condition is always satisfied and relate to   Appendix \ref{E} for the study of the solution when the mentioned condition is not met.
We observe that the probability $p_{ij}(x_i,x_j)$ is given by
\begin{equation}
p_{ij}(x_i,x_j)=\frac{G_{ij}(x_i,x_j)Z_{i|j}(x_i)Z_{j|i}(x_j)}{{Z}Z_{i|j}(x_i)Z_{j|i}(x_j)-F_{ij}(x_i,x_j)}.
\end{equation}

The stationary solution $(\ref{A16})$  can be also written as 
\begin{equation}
b_{ij}(x_i,x_j)=\frac{G_{ij}(x_i,x_j)/F_{ij}(x_i,x_j)}{e^{\beta [\epsilon_{ij}(x_i,x_j)-\mu]}-1}
\label{Auno}
\end{equation} 
valid for $Z e^{\epsilon_{ij}(x_i,x_j)}-1>0$, where $\epsilon_{ij}(x_i,x_j)$ and $\mu$ are defined as
\begin{eqnarray}
\epsilon_{ij}(x_i,x_j)-\mu&=&\frac{1}{\beta}\ln\left[\frac{ZZ_{i|j}(x_i)Z_{j|i}(x_j)}{F_{ij}(x_i,x_j)}\right].\nonumber \\
\label{Aeij}
\end{eqnarray}
Using the relation $(\ref{Am})$ and the Eq.s $(\ref{A16})-(\ref{Auno})$, we derive the marginal  probability  $p_{ij}(x_i,x_j)$, of linked pairs of loci $(i,j)$,
\begin{eqnarray}
p_{ij}(x_i,x_j)&=&b_{ij}(x_i,x_j)Z_{i|j}(x_i)Z_{j|i}(x_j)\nonumber\\
&=&\frac{1}{Z}G_{ij}(x_i,x_j)\{1+n_B[\epsilon_{ij}(x_i,x_j)]\}
\label{pij0}
\end{eqnarray}
with $n_B[\epsilon_{ij}(x_i,x_j)]$ indicating the Bose distribution  \cite{Huang} associated with "energy level" $\epsilon_{ij}(x_i,x_j)$, i.e.
\begin{equation}
n_B[\epsilon_{ij}(x_i,x_j)]=\frac{1}{e^{\beta [\epsilon_{ij}(x_i,x_j)-\mu]}-1}.
\label{ABose}
\end{equation}
Since the distribution of $P(\{x\})$ is normalized, $\mu$ must satisfy, for every pair of linked loci $(i,j)$, the normalization  condition
\begin{eqnarray}
1&=&\sum_{x_ix_j}p_{ij}(x_i,x_j).
\label{Anorm}
\end{eqnarray}
Using  the expression $(\ref{pij0})$ for the marginal distribution $p_{ij}(x_i,x_j)$ we arrive at a set of  equations,
\begin{equation}
1=\frac{1}{Z}\sum_{x_i,x_j}G_{ij}(x_i,x_j)+\frac{1}{Z}\sum_{x_i,x_j}G_{ij}(x_i,x_j)n_B[\epsilon_{ij}(x_i,x_j)]
\label{AZl}
\end{equation}
valid for every pair of linked loci $(i,j)$.
Summing Eq. $(\ref{AZl})$ over every pair of linked loci $(i,j)$ we obtain
\begin{equation}
1=\frac{1}{Z}\int d\epsilon g_{\beta}(\epsilon)[1+n_B(\epsilon)]
\label{AZ0}
\end{equation}
where $n_B(\epsilon)$ is the Bose-Einstein distribution $(\ref{ABose})$ and $g_{\beta}(\epsilon)$
is given by
\begin{eqnarray}
g_{\beta}(\epsilon)&=&\lim_{\Delta \epsilon \rightarrow 0}\frac{1}{\Delta\epsilon}\frac{1}{L}\sum_{i,j\in \partial i}\sum_{x_i,x_j}G_{i,j}(x_i,x_j)\nonumber \\
&&\times\chi_{\Delta \epsilon}(\epsilon_{ij}(x_i,x_j)-\epsilon)
\label{Agbetae}
\end{eqnarray}
with $\chi_\delta(x)=1$ if $|x|\leq \delta$ and $\chi_{\delta}(x)=0$ otherwise.
Therefore the average fitness of the evolving population can be expressed in terms of an integral over a Bose-Einstein distribution with the "energy levels" to be determined self-consistently by the cavity method.

\section{Condensation transition}
\label{E}
An unexpected and new phenomenon can occur in this evolutionary process.
Due to the fact that the joint probability of pairs of allelic states can be expressed in terms of a Bose-Einstein distribution we can predict that in this evolutionary dynamics a condensation in the same universality class as the Bose-Einstein condensation might occur.
In a quantum Bose gas \cite{Huang}, a Bose-Einstein condensation is a phase transition at a critical value of the inverse temperature $\beta_c$ such that for $\beta>\beta_c$ a finite fraction of the total number of particles is found in the ground state.
The equivalent of this phase transition for the evolutionary dynamics described in this paper occurs when  a finite fraction of pairs of loci gets fixed in given allelic configurations.
Therefore, when  this phenomenon occurs in an evolving diploid population, the number of polymorphic  pairs of    loci  is reduced by  a  finite fraction.

Let us  consider the case in which a pair of loci is fixed in the population,
i.e.
\begin{equation}
p_{ij}(x_i,x_j)=\delta(x_i,x_i^{\star})\delta(x_j,x_j^{\star}).
\label{Afix}
\end{equation} 
We want to prove that this condition is equivalent to the condition
\begin{equation}
\epsilon_{ij}(x_i^{\star},x_j^{\star})=\mu.
\label{Acm}
\end{equation}
Given Eq. $(\ref{fix})$, we derive from the definition $(\ref{Adef})$ that the function $G_{ij}(x_i,x_j)$ is a constant and equal to zero, i.e. $G_{ij}(x_i,x_j)=0\, , \forall (x_i,x_j)$.  This result is evident if we observe that  the only  contributions to $G_{ij}(x_i,x_j)$, defined in $(\ref{Adef})$, are given by different pairs of allelic states $(x_i,x_j)$ in the two parental gametes. Since we have assumed that all gametes have the same pair of allelic states in  the genetic loci $(i,j)$, $G_{ij}(x_i,x_j)=0\, \forall (x_i,x_j)$.
Inserting this result in  Eq. $(\ref{Aself})$ we get the following relations
\begin{eqnarray}
b_{ij}(x_i,x_j)=0 \,\mbox{if} \,(x_i,x_j)\neq (x_i^{\star},x_j^{\star})\nonumber \\
\epsilon_{ij}(x_i,x_j)=\mu \,\mbox{if}\,(x_i,x_j)=(x_i^{\star},x_j^{\star}).
\end{eqnarray}
Similarly it is easy to prove that Eq. $(\ref{Acm})$ implies Eq. $(\ref{Afix})$.
Therefore, if we want to describe fixed genetic loci,  we have to modify Eq.  $(\ref{pij0})$, for the marginal probability  $p_{ij}(x_i,x_j)$ according to the following expression,
\begin{eqnarray}
p_{ij}(x_i,x_j)=\left\{\begin{array}{lcl}
1\,\mbox{if}\,\epsilon_{ij}(x_i,x_j)=\mu\nonumber \\\frac{1}{Z}{G_{ij}(x_i,x_j)}\{1+n_B[\epsilon_{ij}(x_i,x_j)]\}\,\mbox{otherwise}\\
\end{array}\right.
\label{Afre}
\end{eqnarray}
Accordingly,  expressions $(\ref{A16})$ and $(\ref{Auno})$ for $b_{ij}(x_i,x_j)$ have to be modified in order to take into account the possibility that a pair of loci gets fixed. Therefore we have
\begin{eqnarray}
b_{ij}(x_i,x_j)=\left\{\begin{array}{lcc}\frac{G_{ij}(x_i,x_j)}{F_{ij}(x_i,x_j)}n_B [\epsilon_{ij}(x_i,x_j)] \, &\mbox{if} & \epsilon_{ij}(x_i,x_j)>\mu \\
\left[Z_{i|j}(x_i) Z_{j|i}(x_j)\right]^{-1} \, &\mbox{if} &\epsilon_{ij}(x_i,x_j)=\mu.\end{array}\right..\nonumber
\label{Abijc}
\end{eqnarray}
The set of equations $(\ref{AZl})$ consistent with the normalization condition $(\ref{Anorm})$ is therefore modified  and  takse the  form
\begin{eqnarray}
\label{AZz}
(1-\nu_{ij})&=&\frac{1}{Z}\sum_{x_i,x_j}G_{ij}(x_i,x_j)+  \\
&+&\frac{1}{Z}\sum_{\begin{array}{c}x_i,x_j\\ \epsilon_{ij}(x_i,x_j)>\mu\end{array}}G_{ij}(x_i,x_j)n_B[\epsilon_{ij}(x_i,x_j)]\nonumber
\end{eqnarray}
valid for every pair of linked loci $(i,j)$. In Eq. $(\ref{AZz})$ the matrix elements $\nu_{ij}$ are taken such that  $\nu_{ij}=1$ if a pair of configurations $(x_i^{\star},x_j^{\star})$ exists such that $\epsilon_{ij}(x_i^{\star},x_j^{\star})=\mu$, otherwise we have $\nu_{ij}=0$. 
The study of the normalization equation $(\ref{AZz})$   will define if and when the number of pairs of fixed  loci becomes extensive.
In the presence of a negligible fraction of fixed pairs of loci, averaging $(\ref{AZz})$ over all pairs of links we get the equation $(\ref{AZ0})$.
For the values of the evolutionary pressure for which Eq. $(\ref{AZ0})$ cannot be satisfied, a finite fraction $\nu_0$ of genetic loci is fixed  and the conservation equation $(\ref{AZ0})$ has to be modified according to 
\begin{equation}
(1-\nu_0)=\frac{1}{Z}\int_{\epsilon>\mu} d\epsilon g_{\beta}(\epsilon)\left[ 1+\frac{1}{e^{\beta (\epsilon-\mu_c)}-1}\right]
\end{equation} 
with $g_{\beta}(\epsilon)$ given by $(\ref{Agbetae})$ and $\nu_0$ defined as
\begin{equation}
\nu_0=\frac{1}{2L}\sum_{i,j\partial i}\sum_{x_i,x_j}\chi_0[\epsilon_{ij}(x_i,x_j)-\mu],
\end{equation}
where $\chi_{\delta}(x)=1$ if $|x|\leq\delta$ and $\chi_{\delta}(x)=0$ otherwise.
As a function of the evolutionary pressure, a condensation transition can occur between a  "non condensed phase" in which all the genetic loci are polymorphic, and a "condensed phase" in which only a fraction of the genetic loci is polymorphic.
This phase transition is in the universality class of the Bose-Einstein condensation and it can be compared with other condensation phase transitions in haploid and diploid evolution \cite{Eigen,Peliti,Kingman,Shraiman,Kadanoff}.

\begin{acknowledgments}
The authors thank Paola Ricciardi-Castagnoli for interesting comments and discussions.
\end{acknowledgments}


\begin{thebibliography}{99}

\bibitem{Fisher}
R. A. Fisher, {\em The Genetical Theory of Natural Selection},
 (Clarendon Press, Oxford, 1930).
\bibitem{Kimura}
M. Kimura, J. of Appl. Probab. {\bf 1}, 177 (1964).

 \bibitem{Hirsh}
 G. Sella and A. E. Hirsh,
Proc. Natl. Acad. Sci. USA {\bf 102}, 9541 (2005).
\bibitem{Gerland}
U. Gerland, J. D. Moroz and T. Hwa, 
Proc. Natl. Acad. Sci. USA {\bf 99}, 12015 (2002).
\bibitem{Monod}
J. Monod, {\em Chance and Necessity: An Essay on the Natural Philosophy of Modern Biology}, (William Collins Sons \& Co., Glasgow, 1972).

\bibitem{Eigen}
M. Eigen, { Naturwiss.} {\bf 58}, 465 (1971).

\bibitem{Peliti}
S. Franz , L. Peliti,  {J. Phys. A} {\bf 30}, 4481 (1997). 

\bibitem{Nowak}
M. A. Nowak, {\em Evolutionary Dynamics},
(Harvard University Press, Cambridge, MA, 2006).

\bibitem{Kingman}
J. F. C. Kingman,
{ J. Appl. Probab.} {\bf 15}, 1 (1978).

\bibitem{Bose}
G. Bianconi and  A. L. Barab\'asi, 
{ Phys. Rev. Lett.} {\bf 86}, 5632 (2001).


\bibitem{Kadanoff}
S. N. Coppersmith, R. D.  Blanck and L. P. Kadanoff,  
Jour. Stat. Phys. {\bf 97}, 1999 (2004). 
\bibitem{Ferretti}
G. Bianconi, L. Ferretti  and S. Franz,   { EPL} {\bf 87}, 28001 (2009).

\bibitem{Shraiman}
R. A.  Neher  and  B. I. Shraiman,  Proc. Natl. Acad. Sci. USA {\bf 106}, 6866 (2009).
 
\bibitem{Genome}
Eric S. Lander et al.  { Nature} { \bf 409}, 860 (2001).
\bibitem{Venter}
J. C. Venter  et al., { Science} {\bf 291}, 1304 (2001).

\bibitem{Barabasi}
A.-L. Barab\'asi and Z. Oltvai,  { Nat. Rev. Genet.} {\bf 5}, 101 (2004). 
 \bibitem{Toroczkai}
E. Ben-Naim, H.  Frauenfelder  and A. Toroczkai,   {\em Complex Networks} Lecture Notes in Physics 650, (Springer-Verlag, 2004).
\bibitem{Sneppen}
K. Sneppen  and G. Zocchi,   {\em Physics in molecular biology},
(Cambridge University Press, Cambridge, England,2005).
\bibitem{Bornholdt}
S. Bornholdt, { Science} {\bf 310}, 449 (2005).

\bibitem{Reka}
R. Albert, {  Jour. of Cell Sci.}
{\bf 118}, 4947 (2005).
\bibitem{Alon}
U. Alon,  {\em  An introduction to system biology: design principles of biological circuits},
(Chapman \& Hall, London, 2007).
\bibitem{Slatkin}
M. Slatkin, {Nature Rev. Genet.} {\bf 9}, 477 (2008).
\bibitem{Kishony}
D. Segr\'e , A. Deluna , G. M. Church and R.   Kishony, { Nature Gen.} {\bf 37}, 77 (2009).
\bibitem{Landscape}
M. Costanzo M et al., Science {\bf 327}, 425 (2010).
\bibitem{Modular}
E. Ravasz, A.L.  Somera, D. A. Mongru, Z. N. Oltvai  and A. L. Barab\'asi, {Science} {\bf 297}, 1551 (2002).
\bibitem{Protein}
H. Jeong, S. P. Mason, A. L. Barab\'asi  and Z. N. Oltvai,
{Nature} (london) {\bf 411}, 41 (2001).
\bibitem{ProteinSK}
S. Maslov  and K. Sneppen,  {Science} {\bf 296}, 910 (2002).
\bibitem{Maritan}
A. Vazquez,  A. Flammini, A. Maritan  and A. Vespignani, { Nature Biotech.} {\bf 21}, 697 (2003).

\bibitem{Metabolic}
H. Jeong, B. Tombor, R. Albert, Z. N. Oltvai and A.-L. Barab\'asi 
{\em Nature} (London) {\bf 407}, 651 (2000).
\bibitem{Blood}
C. I. Jones et al.,  
{ Blood} {\bf 114}, 1406 (2009).
\bibitem{Lander}
D. E. Reich et al., { Nature} (London) {\bf 411},  199 (2001).

\bibitem{Sigmund}
J.  Hofbauer  and K. Sigmund,   {\em Evolutionary Games and Population Dynamics}, (Cambridge University Press, Cambridge, England,1998).

\bibitem{Gillespie}
J. H.  Gillespie,   {\em Population Genetics: A concise Guide}, (John Hopkins University Press, Baltimore, MD, 2004).
\bibitem{Hartl}
D. Hartl  and A. G. Clark,  {\em Principles of population genetics}, (Sinauer Associates Inc. Publisher, Sunderland, 2007).
\bibitem{Sequence}
K. Jain  and J.  Krug,   in {\em Structural Approaches to Sequence Evolution}, eds. U. Bastolla, M. Porto, H.E. Roman, M. Vendruscolo (Springer-Verlag, Berlin, 2007).

\bibitem{Darwin}
C. Darwin,   {\em On the Origin of Species by means of natural selection}, 
(Oxford University Press, New York,1859).

\bibitem{Parisi} 
M. M\'ezard  and G. Parisi, Eur. Phys. Jour. {\bf 20}, 217 (2001).
\bibitem{Yedidia}
J. S. Yedidia, W. T. Freeman  and Y. Weiss,
in: {\em Exploring artificial Intelligence in the New Millennium} (Science and Technology Books,2003).
\bibitem{Weigt} 
A. Hartmann  and M. Weigt,   {\it Phase Transitions in Combinatorial Oprimization Problems},
(Wiley-VCH, Weinheim, 2005). 
\bibitem{Mezard}
M. M\'ezard  and A. Montanari,   {\em Information, Physics and Computation},
(Oxford University Press, Oxford, 2009).

\bibitem{Huang}
K. Huang   {\em Statistical Mechanics},
(John Wiley and  Sons,1987).

\bibitem{Quantumgas}
C. J. Pethick  and H. Smith  {\em Bose-Einstein Condensation in Diluted Gases},
(Cambridge University Press, Cambridge, 2001).

\bibitem{MS}
J. Maynard Smith and J. Haigh, Genetics Research {\bf 23}, 23 (1974).

\bibitem{Peliti2}
L. Peliti and U. Bastolla, CR Acad. Sci. III {\bf 317}, 371 (1994).


\bibitem{Moran}
P. A. P. Moran, Ann. Hum. Genet. {\bf 27}, 383
(1964).

\bibitem{Davies}
P. C. W. Davies, BioSystems {\bf 78}, 69 (2004).

\bibitem{Lloyd}
S. Lloyd,   { Nature Physics} {\bf 5}, 164 (2009).




\end{thebibliography}
\end{document}